\documentclass[twocolumn,showpacs,amsmath,aps]{revtex4}
\usepackage{graphicx,color}
\usepackage{bm}

\begin{document}

\title{Transport Signatures of a Majorana Qubit and Read-out-induced Dephasing}

\author{Lupei Qin}
\affiliation{Institut f\"ur Theoretische Festk\"orperphysik, Karlsruhe Institute of Technology (KIT), 76131 Karlsruhe, Germany}
\affiliation{Center for Joint Quantum Studies, School of Science, Tianjin University, 300072, China}
\author{Xin-Qi Li }
\affiliation{Center for Joint Quantum Studies, School of Science, Tianjin University, 300072, China}
\author{Alexander Shnirman}
\affiliation {Institut f\"ur Theorie der Kondensierten Materie, Karlsruhe Institute of Technology (KIT), 76131 Karlsruhe, Germany}
\affiliation {Institut f\"ur Nanotechnologie, Karlsruhe Institute of Technology (KIT), 76021 Karlsruhe, Germany }
\author{Gerd Sch\"on}
\affiliation {Institut f\"ur Theoretische Festk\"orperphysik, Karlsruhe Institute of Technology (KIT), 76131 Karlsruhe, Germany}

\date{\today}

\begin{abstract}

Motivated by recent proposals of Majorana qubits and the read-out of their quantum state we investigate a qubit setup formed by two parallel topological wires shunted by a superconducting bridge. The wires are further coupled to two quantum dots, which are also linked directly, thus creating an interference loop. The transport current through this system shows an interference pattern which distinguishes two basis states of the qubit in a QND measurement. We analyze various properties of the interference current and the read-out process, including the resulting dephasing and relaxation. We also analyze the effects of varying control parameters such as gate voltages on the current. The characteristic dependencies may serve as a  signature of Majorana bound states.

\end{abstract}

%\pacs{71.10.Pm,74.78.Na,74.45.+c}
\maketitle

\section{Introduction}

Majorana bound states (MBSs) in topological superconductors (TSs) have been proposed as candidates for topologically-protected carriers of quantum information~\cite{Kitaev, Lutchin,Oreg,Das08, Alicea, Leijnse, Beenakker}. Much attention has been paid to the properties of single wires, including the proposal how to perform the topolocically-protected adiabatic braiding operation~\cite{Alicea11, Malciu}. However, in order to overcome limitations from parity conservation and to allow performing a universal set of quantum gates it is necessary to consider generalizations, such as composite multi-wire systems~\cite{Stanescu, Aasen, Plugge16}. Such a system, namely a Majorana box qubit was recently described by Plugge {\sl et al.}~\cite{Plugge}. The basic qubit consists of two TS wires shunted via a conventional superconductor. The two qubit states, both with -- say -- even parity, differ in the number of occupied MBSs. The states can be read out by a conductance measurement, when the Majorana qubit is coupled suitably to electron reservoirs.

Here we revisit the system proposed in Ref.~\onlinecite{Plugge}. On one hand, we propose a specific setup, where the Majorana qubit is coupled via quantum dots to reservoirs. On the other hand, we study  its transport properties and dynamics in the frame of a quantum master equation. We do not discuss protocols how to manipulate the qubit, nor do we consider further TS wires needed for this purpose. However, we study in  detail the read-out process made possible by the interference effects in the transport current.
We find that the setup allows performing the measurement in a quantum non-demolishing fashion.
In addition we obtain information about the time scales of the read-out process such as the relaxation and dephasing induced by the process.

For a current to flow states differing in particle number and hence Fermion parity need to be accessed~\cite{Leijnse11}. Although, due to Coulomb-blockade effects, theses states are only weakly populated (and at $T=0$ only virtually as known from cotunneling), they still influence the dynamics of the system in characteristic ways.
Mixing states with different parity leads to decoherence of the Majorana qubit, similar as the so-called quasiparticle poisoning, although at the low temperatures (considered here) no quasiparticles with energies above the superconducting gap are excited. Since the current should be sufficiently strong for the measurement process, it is reasonable to assume that tunneling is the leading source of decoherence. In comparison, we ignore other mechanisms, such as, e.g., quasiparticle poisoning involving excitations above the superconducting gap~\cite{Albrecht}, or those which arise when the MBSs have a non-zero overlap~\cite{Knapp}. When switched on for the read-out the tunneling leads to a rapid initial decay on the scale of the inverse tunneling rate. In addition we find features of telegraph noise, which manifest themselves in a slow final decay.

We also consider several extensions of the  measurement protocol and the transport properties of the composite system when varying parameters and gate voltages. We find signatures of coherent oscillations, either arising from different states of the dots or involving higher-energy states.
As has been argued before for a similar but more basic setup~\cite{ZZLi}, the sensitive dependence of the results can serve as signatures of the presence of Majorana bound states.

In the following section we will present the model of the system and the relation to the qubit. We then formulate the quantum master equation, which we use to analyze the transport properties and dynamics, and we study the use of the setup for a quantum measurement. We determine the time scales of the various stages of this process, incl.~the dephasing induced by the transport current. We finally consider several extensions and generalizations. This includes the dependence on the voltage bias,  the correlations in the measurement current, and the influence of gate voltages driving the system away from optimum symmetry points (``sweet spot'').

\section{Setup and Model}

\subsection{Hamiltonian and Qubit}

\begin{figure}
  \centering
  \includegraphics[scale=0.8]{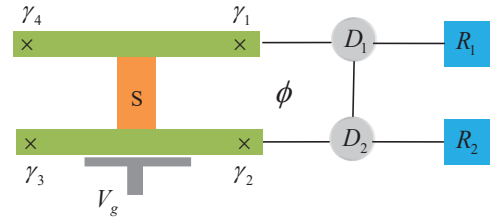}\\
  \caption{Majorana qubit and setup used for current interferometry. Two long TS wires (green) are shunted by a superconducting bridge (orange) to form a floating island hosting four Majoranas $\gamma_{j}$ (crosses).
  For the current measurement we assume that the system is coupled to two quantum dots (grey), which are connected to two independent electron reservoirs (blue). The dots are also coupled directly, thus creating an interference loop with enclosed magnetic flux $\phi$. The gate voltage $V_{g}$ is used to tune the gate charge $n_{g}$.
  }\label{setup}
\end{figure}

We consider the Majorana box qubit (MBQ) displayed in Fig.~\ref{setup}. It is formed by two sufficiently long topological superconductor (TS) nanowires which are shunted by a conventional superconductor S. This creates an electrostatically floating island with charging energy controlled by a gate voltage, but there are no T-junctions of topological superconducting wire segments. The setup hosts four Majorana fermions, $j=1,..., 4$, with $\gamma_{j}^{\phantom\dagger}=\gamma_{j}^\dagger$ and anticommutation relations $\{\gamma_i,\gamma_j\}=2\delta_{i,j}$.
We study long wires, such that the MBSs have negligible overlap and  (approximately) zero energy, which is the origin of their topological protection.
Fermion parity is conserved, $\gamma_{1}\gamma_{2}\gamma_{3}\gamma_{4}=\pm 1$. We concentrate first on the case with even parity (an extension will be discussed later). This leaves still 2 states, which form basis states of the qubit.
Anticipating what will turn out ot be the basis of the measurement process described below, we choose the Pauli operators of the qubit as
$\hat x = i \gamma_{4}\gamma_{1}; \, \hat y = i \gamma_{2}\gamma_{4}; \hat z = i \gamma_{1}\gamma_{2} $.

For the current measurement we assume in the following that the system is coupled to two quantum dots, with energy levels which can be tuned by further gate voltages. They are also coupled directly, thus creating an interference loop with enclosed magnetic flux $\phi$. The dots should be further coupled to electron reservoirs. When turned on, this coupling introduces dissipative processes which will destroy the coherent time evolution. But, as will be shown below, the interference current between the two reservoirs also serves as a measure of the state of the qubit. For a current to flow through the Majorana system we need excited states differing in particle number and hence fermionic parity. But because of Coulomb-blockade effects at low temperature theses states are only vitually/weakly excited.

We model the setup by the Hamiltonian
\begin{equation}\label{HSys}
H_{\rm S}=H_{\rm M}+H_{C}+H_{\rm D}+H_{\rm I}\, .
\end{equation}
Here, $H_{M}=\frac{i}{2}\left(\epsilon_{\rm t}\gamma_{4}\gamma_{1} + \epsilon_{\rm b} \gamma_{2}\gamma_{3}\right)$ is the Hamiltonian of the Majoranas in the top and bottom wire. Ideally the overlap of the Majorana bound states vanishes and  $\epsilon_{\rm t/b}=0$. Hence, this part of the Hamiltonian vanishes.
The next term accounts for the Coulomb charging energy
\begin{equation}\label{HC}
  H_{C}=E_{C}(N-n_{g})^{2} \, .
\end{equation}
It depends on the total charge on the floating island, $N$, which is conserved on the Hamiltonian level but in general varies when a current is flowing. The gate charge $n_g$ depends on the gate voltage $V_g$ and gate capacitance. Also the energy scale $E_C$ depends on the capacitance of the floating island.
There is no need to go into details, but we note that the optimal point for the qubit is a symmetry point with an integer value of $n_g$.  In the following we assume that  $E_C$ is a large energy scale, and all other coupling energies as well as the temparature are much lower.

The quantum dots are assumed to have an even higher on-site interaction energy suppressing double occupancies. In the subspace of empty or singly-occupied dot states the Hamiltonian is given by $H_{\rm D}=\sum_{j=1,2}\epsilon_{j}d^{\dagger}_{j}d_{j}$, with energies $\epsilon_{j}$ wich can be tuned by gate voltages. For definiteness we assume in most of the following that during the read-out process via a current measurement we have $ \epsilon_{1/2}=0$.

The tunneling between the two dots and between each of the dots and the adjacent MBSs are described  by~\cite{Dong11}
\begin{eqnarray}
  H_{\rm I} =& & - \lambda_{0}e^{i\phi}d_{2}^{\dagger} d_{1} - \lambda_{0}^{*}e^{-i\phi}d_{1}^{\dagger}d_{2}  \notag \\
   &+&(\lambda_{1}d_{1}-\lambda^{\ast}_{1}d_{1}^{\dagger})\gamma_{1} +  i(\lambda_{2}d_{2}+\lambda^{\ast}_{2}d_{2}^{\dagger})\gamma_{2} \, .
\end{eqnarray}
The  flux enclosed in the interence loop of the sutup is accounted for by the phase factor $e^{i\phi}$ multiplying the amplitude $\lambda_0$. The asymmetric form of phases of the coupling amplitudes between dots 1 and 2 and the respective Majoranas  is chosen for convenience to produce simple results in the following for real values of $\lambda_{0/1/2}$.

To proceed it is useful to switch to a representation where  two Majorana fermions are combined to form a regular fermion. In anticipation of what will turn out to be relevant in the following we introduce $ f_{\rm R}^{\dag}=(\gamma_1-i\gamma_2)/2$.
Note that the fermion created by $f_{\rm R}^{\dag}$ does not correspond to a pair of Majoranas in the top or bottom TS but rather to a pair on the right side of the setup. Similarly we can define a fermi operator $f_{\rm L}^{\dag}$ by combining the two Majoranas on the left side, but with the assumptions made  it does not enter the Hamiltonian. Thus we find the following Hamiltonian  for the coupling between the dots and the Majorana qubit
\begin{eqnarray}
  H_{\rm I} &=& - \lambda_{0}e^{i\phi}d_{2}^{\dagger} d_{1} - \lambda_{0}^{*}e^{-i\phi}d_{1}^{\dagger}d_{2}   \notag \\ \nonumber
   &+&\lambda_{1}d_{1}f_{\rm R}^{\dag}+\lambda_{1} e^{i\theta} d_{1}f_{\rm R}
-\lambda_{1}^{*} e^{-i\theta} d_{1}^{\dag} f_{\rm R}^{\dag}-\lambda_{1}^{*}d_{1}^{\dag} f_{\rm R} \\ \nonumber
   &+&\lambda_{2}f_{\rm R}^{\dag}d_{2}
   - \lambda_{2} e^{i\theta}  f_{\rm R}d_{2}+\lambda_{2}^{*} e^{-i\theta}  f_{\rm R}^{\dag}d_{2}^{\dag}
-\lambda_{2}^{*}f_{\rm R}d_{2}^{\dag} \, . \\
\end{eqnarray}
Here, $e^{i\theta}$ is the operator, in the gauge proposed in Ref.~\cite{vanHeck}, that adds one Cooper pair to the condensate.

The dots are coupled to electronic reservoirs $j=1,2$.
They are assumed to be formed by free electrons with densities of states $\nu_j$ and states occupied according to a Fermi function depending on the respective voltages.  The tunneling processes between the dots and the reservoirs (with creation  operators $c_{j,k}^{\dag}$), following from $H_{\rm T}=\sum_k (t _{1}d^{\dag}_{1}c_{1,k}+t_{2}d^{\dag}_{2}c_{2,k}+h.c.)$, lead to incoherent transitions with rates
$\Gamma_j = 2\pi \nu_j |t_j|^2$ and consequences to be discussed further below. Throughout this paper we will assume that both rates are equal, $\Gamma_1=\Gamma_2= \Gamma$.

\subsection{Basis and Eigenstates}
We proceed assuming that the ground states of the Majorana qubit have even total parity, and that the total number of charges, $N$,  is even.
In this case the two degenerate ground states  are
\begin{equation}\label{ground states}
|``0_{\rm L}\mbox{"}\rangle =| 0_{\rm L}, 0_{\rm R},  N, n_{0} \rangle \; \; \mbox{and} \; \;
|``1_{\rm L}\mbox{"}\rangle = |1_{\rm L}, 1_{\rm R},  N,  n_{0}-1 \rangle
\end{equation}
They are also the eigenstates of $\hat z$ as defined above and  basis states of the logical Majorana qubit.
Here, the labels $0/1_{\rm L/R}$ denote the parity of the left/right pair of Majoranas in the wires, and
$N$ counts the total charge on the floating island.
The numbers of Cooper pairs, $n_0$ and $n_0-1$,  are adjusted to yield the same total number of charges $N$ in the two states in spite of the difference by 4 Majoranas \cite{vanHeck, Fu}. Due to the coupling to the dots the number of Majoranas on the right-hand side of the wires and hence the total charge can change. We thus have to consider an extended set of states, which includes odd total parity states,
\begin{align}
\label{setsofstates}
&|0_{\rm L}, 0_{\rm R}, N, \, n_{0} \rangle&   &|1_{\rm L}, 1_{\rm R}, N, n_{0}-1 \rangle&\\ \nonumber
&|0_{\rm L}, 1_{\rm R}, N+1, n_{0} \rangle& &|1_{\rm L}, 0_{\rm R}, N+1, \,  n_{0} \rangle&\\ \nonumber
&| 0_{\rm L}, 1_{\rm R}, N-1, \, n_{0}-1 \rangle& &|1_{\rm L}, 0_{\rm R}, N-1,\,  n_{0}-1 \rangle \, .&
\end{align}
Since we assume that the charging energy scale $E_C$ is large we can ignore states with still larger or smaller numbers of the total charge. Tunneling between the dots and the wires induces transitions between the left 3 states belonging to the $``0_L\mbox{"}$ block, and similarly between the right 3 states of the $``1_L\mbox{"}$ block, but it does not -- nor does any other part of the  Hamiltonian --
mix states belonging to different blocks. Note that the parity of the left pair of Majoranas is fixed within either block. This can be seen as the origin of the quantum non-demolition character of the measurement process which we will encounter below.

The above mentioned strong on-site repulsion restricts each one of  the two quantum dots to be empty or singly occupied. We thus have 4 dot basis states $|0,0\rangle, |0,1\rangle, |1,0\rangle, |1,1\rangle$. Hence, the total set of basis states to be considered are 12 product states formed by  the 4 dot states and  the 3 states of the Majorana system from the block $``0_L\mbox{"}$ with even parity on the left side of the system, and similar 12 states involving the block $``1_L\mbox{"}$ with odd parity on the left side of the system.

\begin{figure}
  \centering
  \includegraphics[scale=0.6]{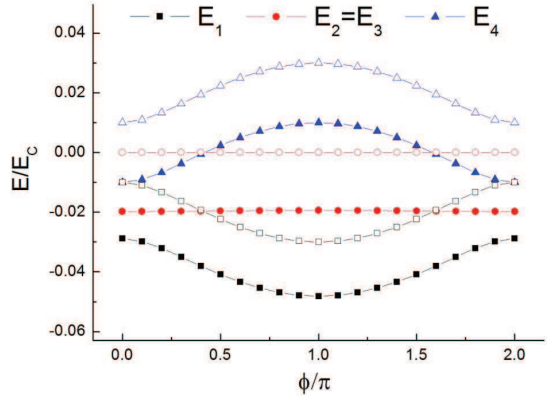}\\
  \caption{The four lowest eigenenergies $E_{1,...4}$ of the coupled Majorana qubit--quantum dot system versus the enclosed flux (normalized to $2\pi$). Only the results for the  $``0_L\mbox{"}$ block are shown. For
  the plot we assumed that  all coupling amplitudes $\lambda_{0/1/2}$ are real.
  For $\phi=0$ the lowest/highest (shown) eigenenergies  (with labels 1 and 4) correspond to dot states $\left(|0,1\rangle \pm |1,0\rangle\right)/\sqrt{2}$, whereas the degenerate middle levels (with labels 2 and 3) are formed by the states $|0,0\rangle$ and $|1,1\rangle$. The energies corresponding to the  $``1_L\mbox{"}$ block have the same $\phi$-dependence but are shifted by $\pi$. The parameters are
 $\lambda_0=0.01 E_C$ and $\lambda_1=\lambda_2 =0.1 E_C$. Open symbols show results obtained from the effective low-energy Hamiltonian (\ref{low-energy}).
  }
  \label{eigenenergies}
\end{figure}

The eigenenergies and eigenstates are easily found. They depend on the enclosed flux as illustrated for the four lowest-energy states in Fig.~2 (with flux normalized to $2\pi$). The results for the two different parity blocks are $\pi$-shifted relative to each other.
The energies can be found analytically. For real $\lambda_0>0$ (i.e., for $\phi=0$ in the figure) and real $\lambda_1=\lambda_2$ the expressions are simple enough to be presented here. We find that one of the eigenenergies is exactly $E_1=-\lambda_0$ and the corresponding eigenstate is a product state of the Majorana qubit ground state and the dot state $|1, 0\rangle + |0, 1\rangle$,  as if the two subsystems would not be coupled.
The next two low-energy states are degenerate with
$E_{2,3}=\frac{1}{2} \left( E_C-\lambda_0-\sqrt{(E_C-\lambda_0)^2+8\lambda_1^2} \right)$
and the 4th state has
$E_4= \frac{1}{2} \left( E_C+\lambda_0-\sqrt{(E_C-\lambda_0)^2+16\lambda_1^2} \right)$.
Similarly for $\phi=\pi$ one of the eigenvalues is exactly $E_4=+\lambda_0$.

In the limit where the tunneling amplitudes
are small compared to the energies of the higher charged states, $|\lambda_\alpha| \ll E_C$, the low-energy eigenvalues can also be obtained -- except for an overall shift by  $-2|\lambda_1\lambda_2|/E_C$ --  from the approximate low-energy Hamiltonian
\begin{equation}\label{low-energy}
H_{\rm eff} = -\left(\lambda_0e^{i\phi} - \hat z \, 2 \frac{\lambda_1 \lambda_2^*}{E_C}\right) d_2^{\dagger}d_1 + {\rm h.c.}
\end{equation}
I.e., the considered system approximately reduces to a double-dot system (with qubit-state-dependent coupling). A similar-looking low-energy Hamiltonian was suggested in Ref.~\onlinecite{Plugge} with operators $d_{1,2}$ referring to two reservoirs, whereas here the dots are part of the coherent quantum system. The factor 2 arises due to two channels for transitions via the two appropriate high-energy states in the blocks (\ref{setsofstates}).

In Fig.~\ref{eigenenergies} we compare the eigenenergies of the effective model and the exact ones. Also in the discussions below we will frequently compare the full model to the low-energy Hamiltonian. The latter reproduces most results with sufficient accuracy, but some details depend on properties of the higher-energy states.

\subsection{Quantum Master Equation}

We next study the properties of  the Majorana qubit--quantum dot system, when it is coupled to two fermionic reservoirs, labelled by $j=1,2$, with chemical potentials $\mu_j$, which can be adjusted to drive a current through the composite system. The reservoirs are assumed to be in thermal equilibrium and their electronic degrees of freedom are traced out. For weak enough coupling the resulting quantum master equation for the system of interest then takes the form~\cite{Li05, Li07, Li06, Jin15, Karlewski1, Karlewski2}
\begin{eqnarray}\label{master equation}
  \dot{\rho}&=&\mathcal{L}\rho   \\
\nonumber &=& \mathcal{L}_{\rm S}\rho -\frac{1}{2}\sum_{j=1,\,2}\{[d_{j}^{\dag},\, D_j^{(-)}\rho
  - \rho D_j^{(+)}]   +h.c.\}\,
\end{eqnarray}
Here, $\mathcal{L}$ is the (total) Liouvillean superoperator which accounts for both the coherent evolution due to the system Hamiltonian,  $\mathcal{L}_{\rm S}\rho\equiv -i[H_{\rm S},\, \rho]$, but also for the dissipation due to the tunneling between the two reservoirs and the adjacent dots.

The dissipative term is assumed to be of the Lindblad form, involving the superoperators
\begin{equation}\label{}
D_j^{(\pm)}
=\int_{-\infty}^{\infty}\,dt\,C_{j}^{(\pm)}(t)e^{\pm i H_{\rm S}t}d_je^{\mp i H_{\rm S}t} \, .
\end{equation}
They depend on the spectral functions of the reservoirs, i.e.,  the thermal averages $C_{j}^{(+)}(t)=\sum_k |t_{j}|^{2} \langle  c^{\dag}_{j,k}(t)c_{j,k}(0)\rangle$ and $C_{j}^{(-)}(t)=\sum_k |t_{j}|^{2} \langle  c_{j,k}(t)c^{\dag}_{j,k}(0)\rangle$.
With the usual approximations for the electrodes the correlation functions become $C_{j}^{(\pm)}(t)=|t_{j}|^{2}\sum_{k}e^{\pm i \epsilon_{k}t}f_{j}^{\pm}(\epsilon_{k})$, where $f_{j}^{+}(\epsilon_{k})=f_{j}(\epsilon_{k})$ is the Fermi distribution   of  reservoir $j$, and $f_{j}^{-}(\epsilon_{k})=1-f_{j}(\epsilon_{k})$. The distribution functions depend on the respective electro-chemical potentials $\mu_j$.
Then in the eigenbasis of $H_{\rm S}$ the superoperator becomes $\left(D_j^{(\pm)}\right)_{nm}=\Gamma_{j}\,f_{j}^{\pm}(\hat \omega_{mn})\,\left(d_{j}\right)_{nm}$,
where $\hat \omega_{mn}=E_m-E_n$ is the energy difference between the eigenstates of $H_{\rm S}$.

The transport current between dot 2 and the reservoir attached to it follows from $I_2(t)=\langle \hat{I_2}(t)\rangle={\rm Tr}\{\hat{I_2}\rho(t)\}$ with current operator
\begin{equation}\label{current}
  \hat{I_2}=\frac{1}{2}\left(d_{2}^{\dag}D^{(-)}_{2}-D^{(+)}_{2}d_{2}^{\dag}\right)+{\rm h.c.} \; ,
\end{equation}
and similar for dot 1. Of course, in the steady state both currents are equal,  $I = I_2=-I_1$, and we can drop the index. However, for transient behavior or for the current correlation functions we have to specify the index.

The formal solution of the quantum master equation
\begin{equation}\label{rho(t)}
\rho(t) = e^{\mathcal{L} t}\rho(0)
\end{equation}
is conveniently recast in matrix form. For this purpose the $N \times N$ density matrix is written as an $N^2$-dimensional vector $\vec \rho$, and the superoperator $\mathcal{L}$ as a $N^2 \times N^2$ matrix which we denote be $\hat G$. The resulting equation
\begin{equation}\label{rho-matrix(t)}
\vec \rho(t)=e^{\hat G t}\vec \rho(0)
\end{equation}
 can be easily solved numerically for the 12 states considered (or 24 when we consider both qubit states together).

The matrix  $\hat G$ is interesting in its own right. It has an eigenvalue zero, and the corresponding eigenvector is the stationary-state density matrix $\rho_{\rm stat}$. When we consider both qubit states together, e.g., if we examine qubit decoherence processes, we find two zero-eigenvalues and corresponding stationary density matrices. One or the other is reached depending on whether the initial state is chosen from one or the other block of states.
Of much interest are also further small eigenvalues of $\hat G$ which govern slow relaxation processes. Below we will encounter examples where physical quantities display such a slow decay.
In addition, as expected, $\hat G$ has many eigenvalues of order $\Gamma_{1/2}$ accounting for the relaxation processes induced by the tunneling between the reservoirs and the dots.

\section{Current Measurement and Read-out of the Qubit State}

\subsection{Steady-state Current}

\begin{figure}
  \centering
  \includegraphics[scale=0.6]{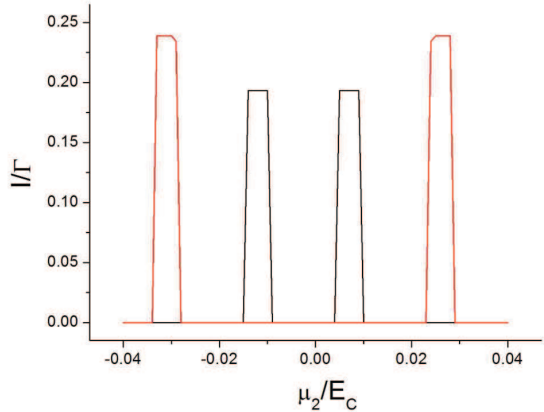}\\
  \caption{
Steady-state current through the system at $T=0$ as a function of the voltage applied to the electrode 2 with a narrow voltage window, $\mu_1-\mu_2= 0.01E_C$. The enclosed flux is $\phi=0$. Results for the two Majorana qubit states are shown by black and red lines.
The other parameters are  $\lambda_0=0.01E_C$, $\lambda_1=\lambda_2 =0.1E_C$, $\Gamma_1=\Gamma_2=\Gamma=0.01E_C$.
  }
  \label{I-V}
\end{figure}

We are now ready to investigate the transport properties of the Majorana qubit--quantum dot system. We control the voltages, i.e. electrochemical potentials of the two reservoirs, concentrating on two scenarios: \\
(i) We can choose a narrow
window $\mu_1-\mu_2$ (for definiteness we assume $\mu_1>\mu_2$). In this case, at low temperatures the current is highly sensitive to the values of $\mu_j$, with a current flowing only when $\mu_1 > \Delta E >\mu_2$, where $\Delta E$ denotes one of the differences in the 4 eigenenergies depicted in Fig.~\ref{eigenenergies}. These energies depend on the Majorana qubit state and on the flux. In Fig. \ref{I-V} we show a typical result.

(ii) Alternatively we can choose a wide window for the difference of the electro-chemical potentials. Specifically we choose  $\mu_1$ and $\mu_2$ such that all transitions between the 4 low-energy states depicted in Fig.~\ref{eigenenergies} are allowed even a $T=0$. This means $\mu_1 > \Delta E > \mu_2$ for all (positive or negative) energy differences. By this choice we avoid switching effects as shown in Fig.~\ref{I-V} which would arise when, upon varying the flux, the energy differences move in or out of the window.
Remarkably, even in this wide-band driving case we find significant interference effects and a dependence of the current on the enclosed flux.

\begin{figure}
  \centering
  \includegraphics[scale=0.6]{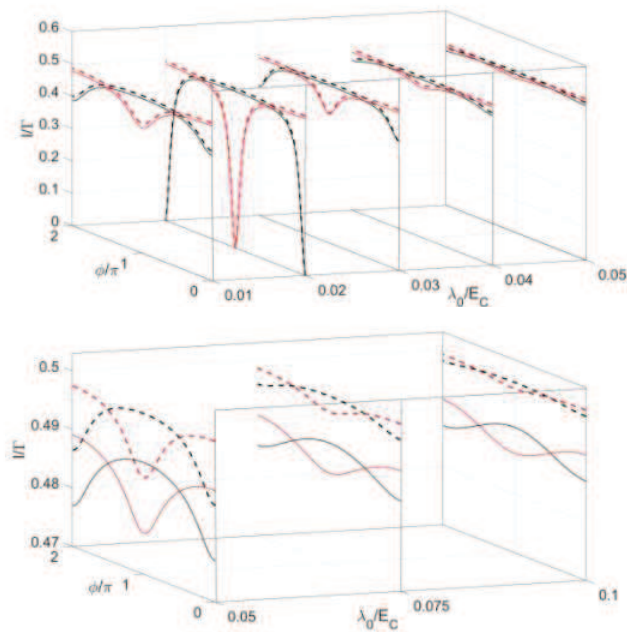}\\
  \caption{Steady-state current  through the system versus enclosed flux for different values of the dot-dot
  coupling $\lambda_0$. The interference effects lead to differences between the results for the two Majorana qubit states (black and red lines). For the plots we assumed that the voltage window of the reservoirs is wide (see the text).   Solid lines are the results obtained for the full 12-state system, dashed lines were obtained from the low-energy Hamiltonian
  with 4 states only.
The parameters are  $\lambda_1=\lambda_2 =0.1E_C$, $\Gamma=0.01E_C$, while $\lambda_0$ varries in the range $0.01 E_C \le \lambda_0 \le 0.1E_C$.
   }
  \label{I-phi}
\end{figure}

In Fig.~\ref{I-phi} we show the steady-state current through the system as obtained for the full 12-state model (see Eq.~\ref{setsofstates}) as a function of the enclosed flux for different values of $\lambda_0$. The results  for the two (dot-dressed) Majorana qubit states differ by a shift by $\pi$. We also show the current as obtained from the effective low-energy Hamiltonian (\ref{low-energy}).
In the case where $\lambda_0$ is relatively large, we find the sinusoidal $\phi$-dependence  predicted for a similar setup in Ref.~\onlinecite{Plugge}. However in this case the amplitude of the current oscillation is small. The amplitude of the current modulation gets larger with decreasing $\lambda_0$, even ranging between a totally constructive or destructive interference for the 2 qubit states, when $\lambda_0=2\lambda_1\lambda_2/E_C$ and $\phi$ is an integer multiple of $\pi$.

We also note that the difference between the results for the full 12-state model and the 4-state low-energy Hamiltonian. They are most pronounced when $\lambda_0$ is large, although they are approximately of the same magnitude for all the displayed curves. The reason is a non-vanishing occupation of the 8 higher energy states of the full model. Although the temperature of the reservoirs is chosen to be low, $T=0$, the fact that the system is driven with a wide voltage window, makes the effective temperature of the low-energy Hamiltonian 4-state system infinite. Similarly, for the full model the four low energy states are roughly equally populated. However, there are also the 8 high-energy states. The eigenstates are superpositions of the low-energy and the high-energy basis states, and even the low-energy eigenstates have a small, but not exponentially suppressed contribution from the high-energy states and vice versa. As a result, also the high energy states acquire a small but not exponentially suppressed occupation probability. This occupation is missing in the probability that the dot 2 states are occupied, accordingly reducing the current out to the reservoir.

 \begin{figure}
  \centering
  \includegraphics[scale=0.3]{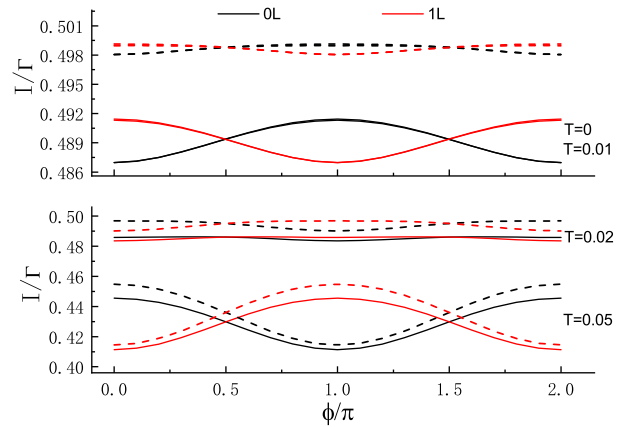}\\
  \caption{Steady-state current  through the system versus enclosed flux for different temperatures. Here $\lambda_0=0.1 E_C$. All other parameters and the choice of lines are the same as in Fig.~\ref{I-phi}.
  }
  \label{I-phi-T}
\end{figure}

In Fig.~\ref{I-phi-T} we show the interference current for different temperatures, concentrating on parameters where at low temperatures the full model and the low-energy effective Hamiltonian produce quantitatively rather different results. At higher temperatures the differences get diminished. Furthermore, we observe a switching of the positions of minima and maxima of the interference current. We found no intuitively simple explanation of this behavior, other than that the interference effects arise due to a subtle balance of transitions.

 Below we will study the dependence of interference current on further parameters. But before doing so, we will investigate the use of the current measurement for a read-out process of the qubit state and investigate the time scales involved.

{\subsection{Dynamics of the Read-out Process}
When the current measurement is performed with the aim to read out the state of the Majorana qubit we should turn on the measurement at some moment, say $t=0$. This can be done, e.g., by tuning gates attached to the quantum dots. Let us assume that the dot energies $\epsilon_{1/2}$ before the measurement are large, then both dots are empty. The measurement starts when we tune them to a lower value, e.g., to   $\epsilon_1=\epsilon_2=0$. An interesting question then is, how does the current evolve in time towards the steady-state value presented above. Fig.~\ref{I(t)} shows such a time evolution of the current $I_2(t)$ from dot 2 to reservoir 2. We show the results for both the full model in black, and those of the effective low-enegry Hamiltonian in red. Obviously the system shows a fast transient behaviour on a time scale given by $\Gamma^{-1}$. The following slow time evolution shows some remarkable properties. For some time we observe also weak oscillations. On one hand, there are high frequency oscillations (displayed only by the full model) with frequency of order $E_C$. They arise since in the transient period states get excited which are superposition of the low and the high-energy states. In addition we observe lower-frequency oscillations, also displayed by the effective low-energy model, with frequency  which coincides with the energy difference between high- and low-energy states of this model $E_4-E_1$.

We further observe that the final relaxation of the current takes place on a slow time scale, slower than $1/\Gamma$, which depends on the values of $\lambda_1$ and $\lambda_2$.
This slow decay process is not observed for the low-energy effective model. The rate for the slow relaxation can also be read off from one of the eigenvalues of the matrix ${\hat G}$. The corresponding eigenstate reveals that in this slow relaxation process the higher charge high-energy states are involved, which are also responsible for the previously mentioned difference in the current obtained for the full and the reduced models.

\begin{figure}
  \centering
  \includegraphics[scale=0.3]{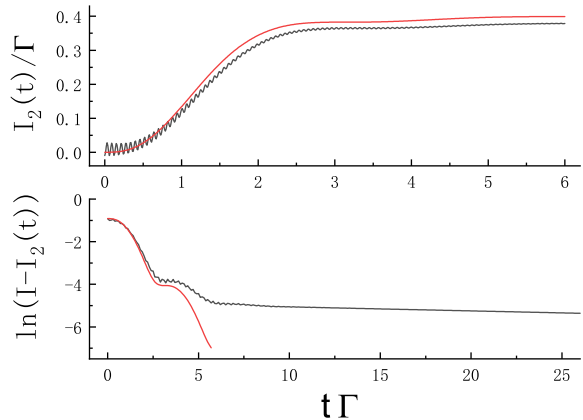}\\
  \caption{a) Current $I_2(t)$ out of dot 2 as a function of the time after the measurement was started. We assume that the initial state is $|0_{\rm L}, 0_{\rm R}, N, \, n_{0} \rangle |0,0\rangle$ with both dot states initially empty.
  The black line is the result obtained for the full 12-state system, the red line arise from the low-energy Hamiltonian
   (\ref{low-energy}).
  b) The deviation of the current $I_2(t)$ from the steady state value $I_2=I$ on a logarithmic scale, displaying for th full model a slow relaxation process.
The  parameters are $\lambda_0=0.01E_C$, the other ones are the same as in Fig.~\ref{I-phi}.  Here and in the following we restrict ourselves to $\phi=0$, except when explicitly mentioned.
    }
  \label{I(t)}
\end{figure}

An important property of the read-out protocol as discussed here is, that it is a quantum non-demolition measurement. The reason, as expected, is that the parity of the left two Majoranas (related to the left Fermion with operators $f_{\rm L}^{(\dagger)}$) is not affected by the tunneling. We reach a stationary current depending on the initial state of the qubit, and this information is not destroyed eventually by the measurement process. When we start from a state from one of the two blocks (\ref{setsofstates}) the system relaxes to a  steady state with the corresponding value of the current. The Liouvillean does not mix the two blocks. When we start with a superposition of states from the two blocks, the relaxation takes place within the blocks, and the current -- as obtained from the master equation, which yields the averages -- is the appropriately weighted average of the results.

\subsection{Majorana Qubit Dephasing}

\begin{figure}
  \centering
  \includegraphics[scale=0.30]{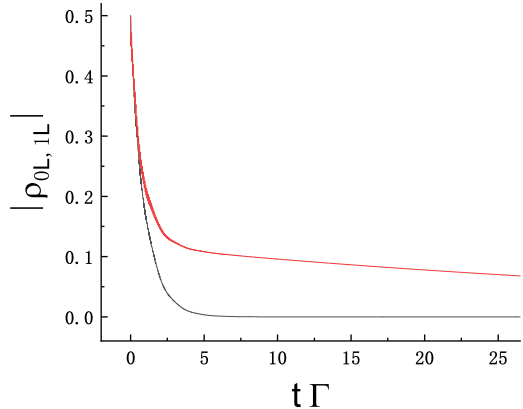}\\
  \caption{Decay of an element of the reduced density matrix $\rho_{0L,1L}={\rm tr}_j\{\rho_{0_L,j;1_L,j}\}$ which is  off-diagonal in the qubit basis but traced over the 4 low-energy dot states. Apart from a rapid initial decay on the time scale $\Gamma^{-1}$ we observe for small $\lambda_0$  a slow dephasing process. Results are shown for $\lambda_0=0.001E_C$ in red and $\lambda_0=0.1E_C$ in black. The remaining parameters are the same as in Fig.~\ref{I-phi}.
 }
  \label{off-diagonal decay}
\end{figure}

The quantum master equation also allows us to investigate the qubit dephasing which is induced by the tunneling of electrons between the reservoirs and the Majorana qubit--quantum dot system.
In Fig.~\ref{off-diagonal decay} we plot an element of the density matrix, which is  off-diagonal  in the two Majorana qubit states. We see a rapid initial decay with rate given by $\Gamma^{-1}$, followed by a slow decay  with rate $\lambda_0^2/\Gamma$, which is visible only in the case when $\lambda_0 \ll \Gamma$. This rate is well known from situations where a qubit is subject to telegraph noise \cite{Paladino,Galperin,Schriefl,Bergli,Matityahu} in the considered limit, where the amplitude of the noise (i.e., the energy shift of the qubit due to the noise) is low compared to rate of switching.  In the present situation there are four low-energy states involved, which changes the details as compared to a single qubit, but one can understand that effectively the two models coincide:
(i) the tunneling of electrons between the reservoirs and the dots with rate $\Gamma$ changes the energy of the system  differing for the two states of the Majorana qubit by an amount of $2 \lambda_0$,
(ii)  in the present model only during half of the time the energies of the two qubit states differ which introduces an extra factor 1/2
 \footnote{When comparing with the results conveniently listed in Ref.~\onlinecite{Matityahu} one should beware of a missing factor 1/2  in the table given there.}.
The result also implies that for  $\lambda_0 =0$ the coherence of the considered superposition would be preserved, i.e., we would have some decoherence-free subspace. However, in this case the interference effect distinguishing the two qubit states vanishes, which makes the process useless for a measurement.

\begin{figure}
  \centering
  \includegraphics[scale=0.30]{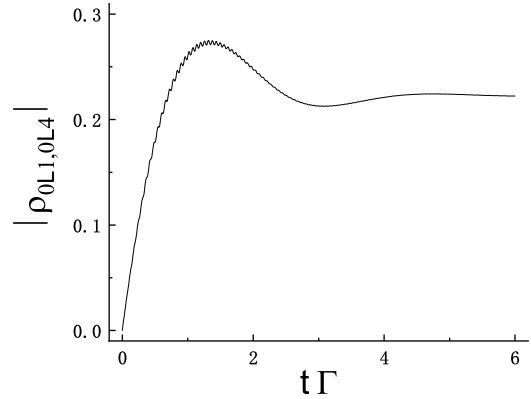}\\
  \caption{Decay of the element $\rho_{0_L1,0_L4}$ of the density matrix which is diagonal in the Majorana qubit states but off-diagonal in the dot eigenstates  (with levels labeled as in Fig.~\ref{eigenenergies}). We note that it does not decay to
zero, since the stationary state is dominated  by the environment.
Results are shown for $\lambda_0=0.01E_C$. The remaining parameters are the same as in Fig.~\ref{I-phi}.
   }
  \label{environment-dominated}
\end{figure}

Next we turn to the issue of  dephasing within the states belonging to  one qubit state, say the $``0_{\rm L}\mbox{"}$ block.
Fig.~\ref{environment-dominated} shows the decay of the element $\rho_{0_L1,0_L 4}(t)$, where the second index in $0_L1$ and $0_L4$ labels one of the four low-energy states of Fig.~\ref{eigenenergies}.
In spite of being off-diagonal in the basis of the Hamiltonian, the element does not decay to zero.
This differs from what one would expect in a perturbative regime. However, the example simply illustrates the difference between Hamiltonian-dominated and  environment-dominated situations~\cite{Zurek}. Under the influence of the environment the eigenbasis of the stationary density matrix differs from that of the Hamiltonian.

\subsection{Current Correlation Function and Power Spectrum}

Next we study the statistical properties of the current, specifically the
symmetrized correlation function of the current $I_2(t)$ out of dot 2 and the power spectrum
\begin{equation}\label{correlation}
S_I(\omega) = \int_{-\infty}^\infty dt \, e^{i\omega t} \langle \{ \delta \hat I_2(t),\delta \hat I_2(0) \}\rangle
\end{equation}
where $ \delta \hat I_2(t) =  \hat I_2(t) - \langle \hat I_2 \rangle$. In the steady state we can rewrite it as
\begin{equation}\label{correlation2}
S_I(\omega) = 2 \mbox{Re} [C_I(\omega)+C_I(-\omega)]
\end{equation}
where
$C_I(\omega) = \int_{0}^\infty dt e^{i \omega t} C_I(t)$
with $C_I(t)= \langle \delta \hat I_2(t)\delta \hat I_2(0) \rangle$.
The latter quantity is conveniently evaluated by using the quantum regression theorem~\cite{Gardiner,Jin15}
\begin{equation}\label{correlation3}
C_I(t) =  \mbox{Tr}\left[\hat I_2  e^{\mathcal{L}t} \hat I_2 \rho_{\rm stat}\right]
 -\langle\hat I_2\rangle^2 \, ,
\end{equation}
where $\rho_{\rm stat}$ denotes the steady state density matrix.

\begin{figure}
  \centering
  \includegraphics[scale=0.3]{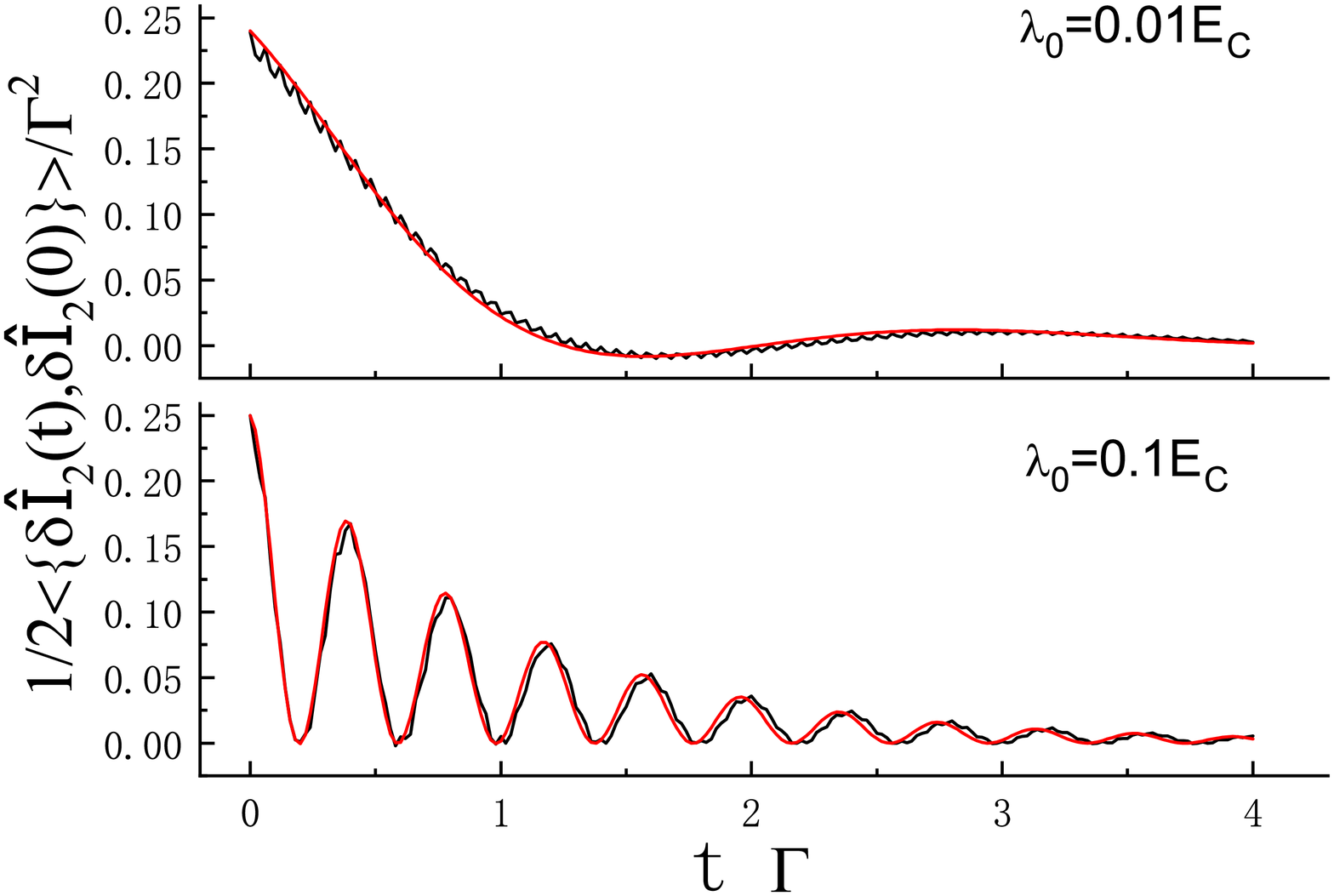}\\
  \caption{Current correlation function $1/2 \langle \{\delta \hat I_2 (t), \delta \hat I_2 (0)\} \rangle$ as a function of time.
 For the upper panel we chose $\lambda_0=0.01E_C$, for the lower panel $\lambda_0=0.1E_C$. The remaining  parameters are the same as in Fig.~\ref{I-phi}.
 Black lines represent the results of the full Majorana qubit--double dot system, red lines are those obtaind with the  effective low-energy Hamiltonian.
}
  \label{fig-correlation}
\end{figure}

\begin{figure}
  \centering
  \includegraphics[scale=0.6]{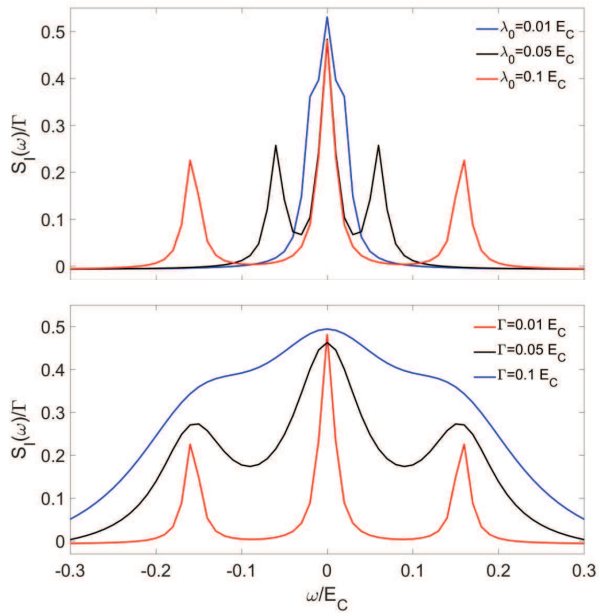}\\
  \caption{Power spectrum $S_I(\omega)$
  (a)  for $\Gamma=0.01 E_C$ and different values of $\lambda_0$,
   (b) for $\lambda_0 =0.1 E_C$ and different values of $\Gamma$.
   The remainig  parameters are the same as in Fig.~\ref{I-phi}.
}
  \label{fig-power}
\end{figure}

In Fig.~\ref{fig-correlation} we show the resulting current correlation function $\mbox{Re}\,  C_I(t)= 1/2 \langle \{\delta \hat I_2 (t), \delta \hat I_2 (0)\} \rangle$. We compare the results for the full 12-state model and the effective low-energy Hamiltonian.
The correlation function displays several of
the characteristic features which we have discussed already above. These are: (i) a decay on the time scale $1/\Gamma$; (ii)  a slow coherent oscillation with frequency which is given by the energy difference $\Delta E  =E_4-E_1$ between the highest and lowest of the four low-energy states. They are strongly pronounced in the lower panel, where this energy difference is larger than $\Gamma$. For the parameters of the upper panel the period of the oscillation exceeds $1/\Gamma$, and the oscillation is hardly visible;  (iii) a high-frequency oscillation, with frequency of order $E_C$ due to the higher energy states. These fast oscillations are best visible when the low-frequency oscillations are overdamped. Of course they can only be seen for the full model.

 In Fig.~\ref{fig-power} we show the resulting power spectrum $S_I(\omega)$ for different values of the tunneling
 rate $\Gamma$. The Fano factor $F \equiv S_I(\omega=0)/2I$ is smaller than 1, as expected for the tunneling of Fermions~\cite{Aghassi}.

\section{Extensions}
\subsection{Varying the Gate Charge $n_{g}$}

\begin{figure}
  \centering
  \includegraphics[scale=0.5]{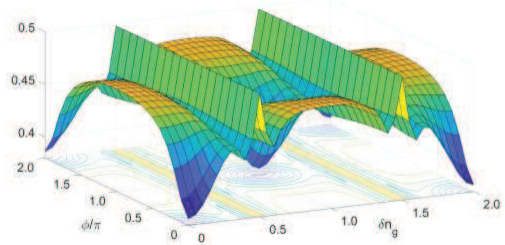}\\
  \caption{Current through the system versus the enclosed flux and the gate charge $n_{g}$.
The remaining  parameters are  the same as in Fig.~\ref{I-phi}.
}
  \label{ng-dependence}
\end{figure}

Extending the analysis presented so far we study the interference current when we vary the gate charge  $n_{g}$. For definiteness we start from the $``0_{\rm L}\mbox{"}$-block of states with even parity on the left side of the two-TS-wire setup. In the situation considered so far the gate voltages attached to the Majorana system where tuned to make $n_g$ an even integer; the state with minimum charging energy had an equally large total charge $N$, and all this was carried by Cooper pairs in the superconducting wire, $N=n_g=2n_0$. The two most important excited states  -- necessary for transport -- as presented in Eq.~(\ref{setsofstates}) had total charge $N+1$ and $N-1$ with charging energy equal to $E_C$ for both. We now allow  $n_{g}=2n_{0}+\delta n_g$ to deviate from an  even number. Accordingly, within the range $\delta n_g\in[0, 1]$, we have to extend the set of states from what is written in Eq.~(\ref{setsofstates}) to
\begin{align}
\label{extendedsetofstates}
&|0_{\rm L}, 0_{\rm R}, N, \, n_{0} \rangle & ,  \; &E(N)=E_{c}\,\delta n_g^{2}&\\ \nonumber
&|0_{\rm L}, 1_{\rm R}, N+1, n_{0} \rangle & , \; &E(N+1)=E_{c}\,(1-\delta n_g)^{2}&\\ \nonumber
&| 0_{\rm L}, 1_{\rm R}, N-1, \, n_{0}-1 \rangle &, \; &E(N-1)=E_{c}\,(1+\delta n_g)^{2}&\\ \nonumber
&|0_{\rm L}, 0_{\rm R}, N+2, n_{0}+1 \rangle &, \; &E(N+2)=E_{c}\,(2-\delta n_g)^{2}& \, .
\end{align}
Here we also indicated the corresponding values of the charging energy. Results for larger values of  $n_g$  follow from symmetry considerations and the periodicity with period 2.

With this extended set of states the simulation can be performed as before. The resulting current for a driven system is displayed in Fig.~\ref{ng-dependence}. For the plot we assumed that the voltage window of the reservoirs is large enough to allow transitions between all  lowest energy eigenstates of the Majorana qubit--quantum dots system. These are 4 states when $n_g$ is close to an integer, but 8 when $n_g$ is close to a half-integer value. Apart from the obvious periodicity when increasing  $n_g$ by 2, we observe the following properties: (i) The current and the interference pattern depends only  very weakly on $n_g$ as long as it is close to an integer. (ii) There is a crossover from
 the behavior corresponding to the $``0_{\rm L}\mbox{"}$-block to that of the $``1_{\rm L}\mbox{"}$-block when $n_g$ is increased by 1. (iii) When $n_g$ is near a half-integer the current takes the maximum value, independent of the flux. Considering Fig.~\ref{ng-dependence} we also note that for the current measurement and read-out process of the Majorana qubit it is not crucial to tune $n_g$ to an even integer with high accuracy. Integer values of $n_g$ correspond to so-called sweet spots, where for symmetry reasons deviations from this spot influence the qubit properties only in quadratic order.

\subsection{Varying the Dot Energies $\epsilon_{1/2}$}

\begin{figure}
  \centering
  \includegraphics[scale=0.6]{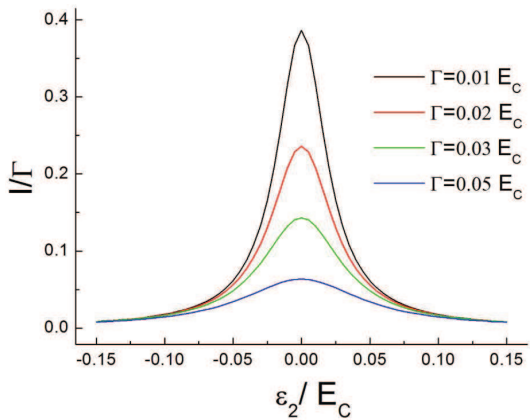}\\
  \caption{Current through the system when the dots are tuned from the symmetry point $\epsilon_j=0$.
The remaining  parameters are  the same as in Fig.~\ref{I-V}.
}
  \label{tune-epsilonj}
\end{figure}
  A similar question is, whether the dot energies, chosen so far as $\epsilon_1=\epsilon_2=0$, have to be tuned exactly to this resonance situation. We therefore allow these energies to deviate from 0. Fig.~\ref{tune-epsilonj} shows that the current decreases, as expected for a out-of resonance situation. For small deviations of $\epsilon_j$ from 0 the effect is quadratic in this deviation, but overall, this dependence appears rather strong.

\section{conclusion}

When analyzing the Majorana qubit--quantum dot system we found  rich physics and
a strong dependence of the transport current on the parameters. As predicted in Ref.~\onlinecite{Plugge} the current between the two reservoirs through the system displays interference effects, which distinguish between the two Majorana qubit states. Remarkably the measurement has a quantum non-demolishing character. This allows reading-out the state of the qubit in a steady state measurement. We studied the time scales characteristic for this process. We also analyzed the decoherence induced by the measurement when the qubit is prepared in a superposition of states.

The electron tunneling into the Majorana qubit--quantum dot system mixes states with different Fermion parity and leads to decoherence.
As expected, the dominant initial relaxation takes place with the rate $\Gamma$ of the tunneling. However, we also find further, previously not discussed decay time scales depending on the coupling energies $\lambda_{0/1/2}$. Specifically, the coherence of the qubit decays with rate $\lambda_0^2/\Gamma$ typical for telegraph noise, which is observable when this rate is smaller than $\Gamma$.

The coherence of Majorana qubits is known to get destroyed by ``quasiparticle poisoning''. Usually this term refers to situations involving excitations above the superconducting gap. In the present work we assumed that they are frozen out, but at higher temperatures they can lead to characteristic ``shadow'' traces in the $I-V$-characteristics when transport and gate voltages are varied~\cite{Albrecht}. It would be interesting to extend the present work to the parameter regime where these properties can be studied, but at this stage we have concentrated on the qubit quantum measurement process.
Somewhat similar to quasiparticle poisoning the electron tunneling considered here leads to decoherence, but we want to recall the specific properties.
The tunneling does not mix the two qubit states, more precisely it does not lead to transitions between the  $``0_L\mbox{"}$ and the  $``1_L\mbox{"}$ blocks. This is the basis for the non-demolishing character of the read-out process by a measurement of the interference current. However the tunneling affects the energies of the two blocks in different ways and thus leads to decoherence, as described with properties similar to the effect of telegraph noise. Quasiparticle poisoning, which could arise for instance because of some normal conducting parts in the system, could lead to transitions between the two blocks and further destroy the coherence.

The analysis presented in the Section ``Extensions'' shows that by choosing gate voltages such that  $n_g$ is integer and $\epsilon_1=\epsilon_2=0$ we have biased the system at a symmetry point, i.e., a sweet spot where the effects of the fluctuations in these parameters and control voltages enter only quadratically. Hence the decoherence induced by these fluctuations is minimized. This fact also supports our strategy to concentrate on the decoherence effects due to tunneling when the transport voltages are turned on and to ignore other sources of fluctuations. On the other hand, the considered system would also be a good test case to study the effect of further noise sources on the dynamics of a multi-level (4 or 12-levels) system. It  differs in detail from a situation with two coupled qubits studied previously~\cite{Governale}.

If some of the features described in this work were detected in an experiment, one could try to fit them quantitatively and thus demonstrate that the model was chosen correctly. The results are very sensitive to the Majorana physics properties. Accordingly the comparison would be a sensitive test of the model and the presence of Majorana bound states.\\

\begin{acknowledgments}

We thank Jinshuang Jin, J.-M. Reiner and M. Marthaler for helpful discussion. Support by
the National Key Research and Development Program of China (No.~2017 YFA0303304), the NNSF of China (No. 11675016),
the Deutsche Forschungsgemeinschaft (Grant No.~SCHO287/7-1), and the DFG-RSF Research Grant SH 81/4-1 is gratefully acknowledged.

\end{acknowledgments}

\end{document}